\documentclass[pdflatex,sn-mathphys-num]{sn-jnl}

\usepackage{graphicx}%
\usepackage{multirow}%
\usepackage{amsmath,amssymb,amsfonts}%
\usepackage{amsthm}%
\usepackage{mathrsfs}%
\usepackage[title]{appendix}%
\usepackage{xcolor}%
\usepackage{textcomp}%
\usepackage{manyfoot}%
\usepackage{booktabs}%
\usepackage{array}%
\usepackage{algorithm}%
\usepackage{algorithmicx}%
\usepackage{algpseudocode}%
\usepackage{listings}%
\usepackage{todonotes}
\usepackage{tcolorbox}
\newtcolorbox{rqanswer}[1]{colback=black!4, colframe=black!60,
  boxrule=0.5pt, arc=1.5pt, left=6pt, right=6pt, top=4pt, bottom=4pt,
  title={#1}, fonttitle=\bfseries\small, colbacktitle=black!60,
  coltitle=white}
\usepackage{fullpage}

\theoremstyle{thmstyleone}%
\theoremstyle{thmstyletwo}%

\theoremstyle{thmstylethree}%

\begin{document}

\title[LLMs for Multi-Level Modelling]{Can LLMs Learn and Apply Multi-Level Modelling Semantics? A First Empirical Study}

\author*[1,2]{\fnm{Yuhong} \sur{Fu}}\email{yuhong.fu@adelaide.edu.au}

\author[3]{\fnm{Weixing} \sur{Zhang}}\email{weixing.zhang@kit.edu}

\author[3]{\fnm{Bowen} \sur{Jiang}}\email{bowen.jiang@kit.edu}

\author[4]{\fnm{Haowei} \sur{Cheng}}\email{haowei.cheng@fuji.waseda.jp}

\author[1,2]{\fnm{Karamjit} \sur{Kaur}}\email{karamjit.kaur@adelaide.edu.au}

\author[1,2]{\fnm{Markus} \sur{Stumptner}}\email{markus.stumptner@adelaide.edu.au}

\affil*[1]{\orgdiv{Industrial AI Research Centre}, \orgname{Adelaide University}, \orgaddress{
\city{Mawson Lakes},
\country{Australia}}}

\affil[2]{\orgname{Future Energy Exports Cooperative Research Centre}}

\affil[3]{\orgdiv{SDQ group}, \orgname{Karlsruhe Institute of Technology}, \orgaddress{
\city{Karlsruhe},
\country{Germany}}}

\affil[4]{\orgname{Waseda University}, \orgaddress{
\city{Tokyo},
\country{Japan}}}

\abstract{Industry 5.0 emphasises human-centric industrial system design, which places additional demands on modelling tools. Compared with traditional two-level modelling, multi-level modelling (MLM) can directly represent three or more abstraction levels, but this expressiveness comes at the cost of more complex semantic constraints, and model correctness depends on these constraints being understood and followed. Large Language Models (LLMs) have been increasingly studied in model-driven engineering, but the resulting evidence rests entirely on two-level modelling tasks, and whether it generalises to MLM, whose semantics differ in kind, remains untested. This paper presents the first empirical study of this question. We have three commercial LLMs (GPT-5.4, Claude Opus 4.6, and Gemini 3.1 Pro) generate multi-level models for the MULTI Warehouse Challenge scenario in the SLICER language under six prompting strategies, with the full language semantic documentation provided in context, yielding 90 generated models in total, which we compare against a manually validated reference model using fourteen metrics. All but one of the 90 models are readable, valid SLICER models, showing that syntactic correctness is within reach; semantic correctness, however, is only partially achieved, with Instantiation/Specialisation Correctness ranging from 52\% to 79\%. The models reproduce content stated explicitly in the task text, but rarely complete structure and constraints that the text implies without stating. Different prompting strategies trade off precision against completeness, and self-checking functions mainly as a rule checker rather than reliably moving generated models closer to the reference design. Among the three LLMs evaluated, Claude shows the most balanced profile in this benchmark. These results provide a structured understanding of the boundaries of current LLM capability for MLM and inform the design of human-centred, AI-assisted modelling workflows for Industry 5.0.}


\maketitle

\section{Introduction}\label{sec:Intro}
In mainstream software engineering practice, models are typically organised into two levels. One level defines the types (for example, a UML class diagram), and the other level defines the instances (for example, an object diagram)~\cite{schmidt2006model}. This class/object paradigm forms the basis of UML. The classic four-layer metamodel architecture of the UML standard (M0 through M3, from runtime objects to models, metamodels, and finally meta-metamodels) appears to consist of four layers, but is essentially the same type-instance relationship nested repeatedly at successive levels: a model at M1 is an instance of a metamodel at M2, and a metamodel at M2 is in turn an instance of a meta-metamodel at M3, and so on~\cite{omg2013meta}. In other words, regardless of how many layers the architecture depicts, what it handles between any two adjacent layers is always a standard two-level relationship. Multi-level modelling (MLM) differs from this: it allows three or more abstraction levels to be represented directly within a single model, without relying on this repeated nesting of two-level relationships~\cite{DBLP:conf/uml/AtkinsonK01}.

The two-level paradigm works well in many scenarios, but it becomes clearly inadequate when a domain naturally involves three or more abstraction levels. Consider, for example, an industrial equipment domain. At the highest level, there might be a general concept of ``Equipment Type'' (such as pumps, valves, or compressors). At the next level, specific models of a particular manufacturer's equipment exist. At the lowest level, individual physical items are deployed in a plant. In a traditional two-level setting, the class/instance mechanism can only directly represent two of these levels at a time. Representing all three requires workarounds, such as introducing extra classes, using stereotypes, or employing the powertype pattern~\cite{DBLP:conf/uml/AtkinsonK01}. While these approaches achieve their purpose, they introduce into the model a form of complexity that has nothing to do with the domain itself and arises purely from a mismatch between the modelling technology and the domain structure, which is what the literature calls ``accidental complexity''~\cite{DBLP:journals/sosym/AtkinsonK08}. More importantly, designing and maintaining these workarounds, and ensuring their consistency across multiple levels, itself constitutes an additional cognitive burden for engineers, one that does not come from domain knowledge itself but from the mismatch between modelling tools and domain structure. In the context of Industry 5.0's emphasis on human-centric system design~\cite{breque2021industry5}, this is precisely the kind of burden that should be absorbed by tools rather than shifted onto individual engineers.

MLM was developed precisely to address this specific, identifiable problem~\cite{DBLP:conf/uml/AtkinsonK01}. It is not ``more expressive'' in some vague general sense, but rather allows a model to span an arbitrary number of abstraction levels directly, removing the need for the workarounds required under a two-level framework. This makes it possible to represent naturally multi-level domain structures, such as product specification types, product specifications, and individual instances, directly and concisely~\cite{DBLP:conf/models/KuhneJ23}. In this respect, MLM has already taken a step toward having tools, rather than engineers, absorb this complexity. However, this expressiveness comes at a cost: to support abstraction at an arbitrary number of levels and consistency across levels, MLM must introduce a set of semantic rules considerably more complex than those of two-level modelling in order to constrain model correctness, and if these rules are not correctly understood and followed, the model cannot be genuinely usable. In other words, MLM takes the burden of level mismatch off the engineer's hands, but in doing so creates a new burden at the level of semantic rules, one that similarly demands specialised expertise to manage.

Large language models (LLMs) have already been explored at scale in model-driven engineering (MDE). A systematic mapping study by Zhang et al.~\cite{zhang2026llm4mde} analysed 86 primary studies published between 2022 and early 2026, and an independent systematic literature review by Di Rocco et al.~\cite{di2025use}, covering studies from 2019 to 2024, reached a convergent conclusion despite a differently scoped search. Both show that the application of LLMs in MDE has grown from sporadic attempts into a rapidly expanding research direction, and both find the surveyed modelling artefacts confined to two-level structures such as domain models, metamodels, and Simulink models, with MLM entirely absent.
These studies commonly rely on two capabilities of LLMs: understanding formal specifications and complex semantic rules, and generating structured text, including text in domain-specific languages. If these two capabilities prove reliable, LLMs could in principle take on this burden even further, not only shifting the complexity of level structures from engineer to tool as MLM already does, but also handing over to the LLM the task of correctly understanding and applying the MLM semantic rules themselves, bringing the field more fully in line with the human-centric modelling vision of Industry 5.0. However, across both reviews, the surveyed studies focus almost entirely on two-level modelling tasks, particularly UML-related model generation, completion, validation, and transformation. MLM is entirely absent, with no study addressing it. This paper presents the first attempt to apply LLMs to MLM tasks.

The semantic constraints that MLM introduces beyond UML differ in kind from those of two-level modelling, not merely in number. For example, a single model element can simultaneously play the dual role of type and instance, both defining the structure of lower-level elements and conforming to the constraints of a type at a higher level (the so-called ``deep instantiation'' mechanism)~\cite{DBLP:conf/uml/AtkinsonK01}. The abstraction level to which a model element belongs is no longer statically given but is instead determined dynamically by the relationships between elements~\cite{DBLP:journals/dke/SelwaySMJGS17}, so that every choice of relation affects the consistency of the entire level structure. 
These constraints have no counterpart in two-level modelling. The existing empirical evidence for LLM capability in MDE has been accumulated entirely within a two-level semantic structure, and it remains unknown whether this evidence, and the capability profile it describes, generalises to a setting whose defining semantic mechanisms that structure does not contain at all.
If LLMs cannot reliably learn and apply this semantics, their ability to relieve engineers of modelling complexity in this multi-level setting likewise remains out of reach.

This paper examines this question by conducting a comparative evaluation of multiple commercial LLMs on MLM tasks using the SLICER language~\cite{DBLP:journals/dke/SelwaySMJGS17}. SLICER defines abstraction levels dynamically through seven typed relations, each carrying distinct semantic meaning. We assess the LLMs' ability to generate correct SLICER models from natural language requirements, using a modelling task based on the MULTI Warehouse Challenge~\cite{DBLP:conf/models/KuhneJ23}, and evaluate the generated models across five dimensions: syntactic correctness, semantic correctness, structural completeness, structural similarity, and consistency.

Within this evaluation, we formulate three research questions:

\textbf{RQ1:} \textit{To what extent can LLMs generate multi-level models that are syntactically correct and semantically valid at both the language and domain levels?}
In this study, all LLMs are provided with the full SLICER semantic specification as in-context documentation, as SLICER is a low-resource language with minimal or no presence in pre-training data. \textbf{RQ1} therefore, evaluates the LLMs' ability to learn and apply formal MLM semantics from context, rather than recall from pre-training knowledge. Answering this question establishes a baseline understanding of current LLM capability for MLM tasks.

\textbf{RQ2:} \textit{How do LLM choice and prompting strategy affect the quality of LLM-generated multi-level models?} RQ1 establishes a baseline for LLM performance under a single prompting condition, but this performance may be shaped by two distinct factors: which LLM is used, and how the task is presented to it. RQ2 therefore jointly examines these two factors within the same experimental design, ranging from zero-shot to strategies that incorporate examples and reasoning steps, together with the LLMs themselves, in order to determine whether differences in generation quality mainly stem from differences in model capability, from differences in prompt design, or from an interaction between the two.

\textbf{RQ3:} \textit{What are the common types of semantic errors made by LLMs in MLM model generation?}
RQ1 and RQ2 measure the overall level of generation quality and how it varies, but they do not reveal which specific semantic decisions LLMs get wrong. RQ3 therefore turns to a structured categorisation of failure patterns, examining recurring types of errors in the generation process to provide a more targeted basis for understanding where the models specifically struggle.

Answering these questions across 90 generated models shows that syntactic validity is easily achieved, while semantic correctness remains only partial, and this shortfall concentrates in two patterns that recur regardless of which LLM or prompting strategy is used. First, the generated models reliably reproduce content stated explicitly in the task, but rarely complete structure or constraints that the text implies without stating, indicating that current LLMs behave more as faithful transcribers than as modellers who complete gaps on their own initiative. Second, errors concentrate specifically in the fine-grained relation kinds that distinguish SLICER from two-level modelling, consistent with LLMs collapsing these distinctions back onto the coarser type-instance opposition more common in their training data.
The contributions of this paper are as follows. First, this work addresses a gap that is structural rather than merely topical: MLM has, to date, been entirely absent from the empirical literature on LLM-assisted modelling, and existing evidence of LLM capability in two-level modelling tasks has never been tested against a semantic structure it does not contain. This paper provides that test.
Second, we design a set of metrics that specifically target multi-level semantic correctness, such as Instantiation/Specialisation Correctness and Semantic Constraint Satisfaction Rate, extending prior LLM-for-modelling evaluation approaches, which have largely measured generic syntactic and semantic correctness, to the level-defining semantic distinctions unique to MLM. Third, by identifying precisely where LLM capability ends and human or rule-based oversight becomes necessary, our results clarify the current limits of applying LLMs in human-centric modelling workflows, a concern central to Industry 5.0.

The remainder of this paper is organised as follows. Section~\ref{sec:Background} introduces the necessary background on LLMs and MLM, including the SLICER language. Section~\ref{sec:method} describes the experimental methodology. Section~\ref{sec:Results} presents the evaluation results for the three research questions. Section~\ref{sec:Discussion} discusses the findings and their implications. Section~\ref{sec:related_work} reviews related work. Section~\ref{sec:Conclusion} concludes the paper.

\section{Background}\label{sec:Background}

\subsection{LLM}
\paragraph{LLM Capabilities for Structured Output Generation}
LLMs have shown strong performance across software engineering tasks such as code generation, code summarisation, and test generation~\cite{DBLP:journals/tosem/HouZLYWLLLGW24, DBLP:journals/chinaf/ZhangFXZYSYC26}. For this paper, the relevant capability is structured generation: producing output that conforms to a target notation while preserving the intended domain semantics.
Research has shown that LLMs handle these two types of constraints with different levels of success. A study by Liu et al.~\cite{DBLP:conf/acl/LiuCSZNH0L24} on LLM proficiency in formal languages found that LLMs understand formal language structure reasonably well but struggle more with generation, particularly when the target formalism is less natural-language-like. 

\paragraph{Prompting Strategies}
LLM performance depends strongly on how the task is represented in the prompt~\cite{DBLP:journals/tacl/MizrahiKMDSS24}. Zero-shot prompting provides the LLM with a task description and expects it to generate a response without any examples. This tests the LLM's ability to apply its general knowledge to an unfamiliar task. Few-shot prompting adds one or more input-output examples to the prompt, allowing the LLM to learn the expected format and content from the examples~\cite{DBLP:conf/nips/BrownMRSKDNSSAA20}. Chain-of-thought (CoT) prompting, introduced by Wei et al.~\cite{DBLP:conf/nips/Wei0SBIXCLZ22}, instructs the LLM to reason through intermediate steps before producing a final answer. Kojima et al.~\cite{DBLP:conf/nips/KojimaGRMI22} showed that even a simple addition such as "Let's think step by step" can improve performance on reasoning tasks. Li et al.~\cite{DBLP:journals/tosem/LiLLJ25} proposed Structured Chain-of-Thought (SCoT) prompting for code generation, which uses program structures as intermediate reasoning steps rather than natural language. These strategies vary depending on whether the model receives examples and whether generation is performed in a single pass or through intermediate modelling steps.

Particularly relevant for generating output in domain-specific languages is grammar prompting, introduced by Wang et al.~\cite{DBLP:conf/nips/WangW0CSK23}. Grammar prompting augments few-shot examples with the formal grammar (typically in BNF form) of the target language. During generation, the LLM first identifies relevant grammar rules and then generates output accordingly. This technique was shown to be effective for diverse DSL tasks including semantic parsing, planning, and molecule generation.

\subsection{MLM}
\paragraph{From Two-Level to Multi-Level Modelling}

As outlined in Section~\ref{sec:Intro}, the class/object paradigm can directly represent only two abstraction levels at a time, and capturing domains that naturally span three or more levels forces modellers into workarounds that introduce accidental complexity~\cite{DBLP:journals/sosym/AtkinsonK08}. Atkinson and K"uhne~\cite{DBLP:conf/uml/AtkinsonK01} identified this limitation in their foundational work on multilevel meta-modelling. MLM addresses it by relating each level to the one above through classification and to the one below through instantiation, so that a single model can accommodate any number of levels~\cite{DBLP:journals/tosem/LaraGC14}. Since the early proposals, a number of MLM tools and languages have been developed, including MetaDepth~\cite{DBLP:conf/tools/LaraG10}, Melanee~\cite{DBLP:conf/modellierung/AtkinsonG16}, MultEcore~\cite{DBLP:conf/models/MaciasRS16}, FMMLx~\cite{DBLP:journals/sosym/Frank22}, DeepTelos~\cite{DBLP:conf/er/JeusfeldN16}, and SLICER~\cite{DBLP:journals/dke/SelwaySMJGS17}. Each of these takes a somewhat different approach to implementing multi-level concepts, but all share the goal of supporting more than two ontological levels within a single model.

\paragraph{Core MLM Concepts}
Although MLM approaches vary in their specific mechanisms, several foundational concepts appear across most of them. Understanding these concepts is important because they define the semantic constraints that any correct multi-level model must satisfy, and therefore the constraints that an LLM must respect when generating such models.

The most basic concept is that model elements can simultaneously act as types (defining the structure of their instances) and as instances (conforming to a type defined at a higher level). In potency-based approaches such as Melanee and MetaDepth, this dual role is captured through a mechanism called ``deep instantiation," where an element called a ``clabject" carries a potency value that controls how many levels of instantiation it can participate in~\cite{DBLP:conf/uml/AtkinsonK01}. When an element is instantiated, the potency of its features decreases by one, and features at potency zero can no longer be instantiated further. SLICER takes a different approach, as described in the next subsection.
A second key concept is the distinction between ontological and linguistic classification~\cite{DBLP:journals/tomacs/AtkinsonK02}. Linguistic classification refers to the relationship between a model element and the metamodel concept that defines its syntactic form (for example, every element in a UML model is linguistically an instance of the UML metaclass ``Class" or ``Association"). Ontological classification, on the other hand, refers to the domain-level relationship between a type and its instances (for example, ``Dog" is an ontological instance of ``Species"). Maintaining this distinction is important because confusing the two leads to models that appear well-formed syntactically but are semantically incorrect.
A third concept is level assignment. Every element in a multi-level model resides at a specific abstraction level. The rules governing which level an element belongs to, and which elements can instantiate or specialise which other elements, vary between MLM approaches. In potency-based approaches, levels are typically assigned based on potency values. In SLICER, levels are determined by the relationships connecting model elements, whereas other approaches may use different mechanisms to define or organise levels.

\paragraph{The SLICER Language}
SLICER was first introduced at the ER 2015 conference~\cite{DBLP:conf/er/SelwaySMJGS15} and later formally specified and extended in a journal article in Data and Knowledge Engineering~\cite{DBLP:journals/dke/SelwaySMJGS17}. It has also been applied to industrial-scale data interoperability in the oil and gas domain~\cite{DBLP:journals/sosym/IgamberdievGSS18} and was recently used to address the MULTI 2024 Warehouse Challenge~\cite{DBLP:conf/models/FuSGKS24}.

Unlike potency-based MLM approaches, SLICER does not use clabjects or potency values. Instead, it uses the term ``object" for all model elements and derives abstraction levels from typed relationships between objects. The seven SLICER relation kinds are:
\begin{itemize}
    \item \textbf{InstN} (normal instantiation), which creates a terminal instance without extending the attribute set;
    \item \textbf{InstX} (extended instantiation), which creates an instance that can extend the attribute set and remain instantiable at lower levels;
    \item \textbf{SpecR} (specialisation by refinement), which refines inherited properties without introducing a new modelling level;
    \item \textbf{SpecX} (specialisation by extension), which extends the inherited vocabulary and introduces a new modelling level;
    \item \textbf{Cat} and \textbf{Member}, which represent categorisation and membership; and
    \item \textbf{SbS} (subset by specification), which links an entity type with a specification type at the same level.
\end{itemize}
The choice of relation type is therefore not just syntactic. It determines level placement, whether lower-level instantiation remains possible, and how attributes and constraints propagate.

SLICER was chosen as the target language for this study for three reasons. First, it has a defined textual syntax that is suitable for LLM-based text generation. Second, it captures the essential semantic distinctions of MLM through its typed relations, making it a meaningful test of whether LLMs can understand MLM semantics. Third, SLICER is a relatively specialised language developed at the University of South Australia, meaning it is unlikely to appear extensively in LLM training corpora. This creates a more genuine test of an LLM's ability to understand and apply formal modelling concepts, rather than simply reproducing patterns memoried from training data.

\paragraph{Attributes and Levels}
In SLICER, attributes are the basic mechanism through which domain-specific relationships are specified between objects. When an attribute is defined, it is given a type (another entity in the model) without the need to specify at which level the attribute is to be assigned. Through specialisation relationships, the type of an attribute can be refined to a more specific entity, consistent with the hierarchy, or simply inherited as is.
Upon instantiation, an attribute may be assigned a value (another entity in the model conforming to the defined type) or refine its type consistent with the instantiation hierarchy (for InstX only). Once a value has been assigned to an attribute, it cannot propagate across further instantiation relationships.
General associations in SLICER are represented as attributes of the source object. The attribute name is the label of the association and the target object is the type of the attribute. When the association appears at a level of instantiation below its definition, the target object serves as the value of the attribute, functioning as a link.
Levels in SLICER are dynamically defined through relationships. SpecX, InstX and InstN introduce new levels. SpecR and SbS do not introduce new levels. Level segregation in SLICER means that only relationships between objects at different levels can introduce a new level. Level cohesion means that SbS relationships must connect objects at the same level, and SpecR does not introduce a new level but can potentially be at a different level of specification.

\paragraph{Constraints and Semantic Axioms}
The constraint mechanism of SLICER permits inheritance of constraints across specialisation, instantiation, and membership relationships. A specialised object must satisfy all of the constraints of the more general object. Constraints in SLICER can propagate across multiple levels of instantiation naturally to the point at which they can be evaluated based on whether the attributes involved in the constraint are assigned.
Combined with SbS relations, constraints in SLICER can be defined where the relevant information is known at the specification level and carried down to the instantiation at which they apply in a well-defined way. If a constraint definition is not applicable at a particular level because attributes are not sufficiently instantiated, the constraint propagates to lower levels until evaluation becomes possible.

\section{Methodology}
\label{sec:method}
\begin{figure*}[htbp]
\centering
    \includegraphics[width=\linewidth]{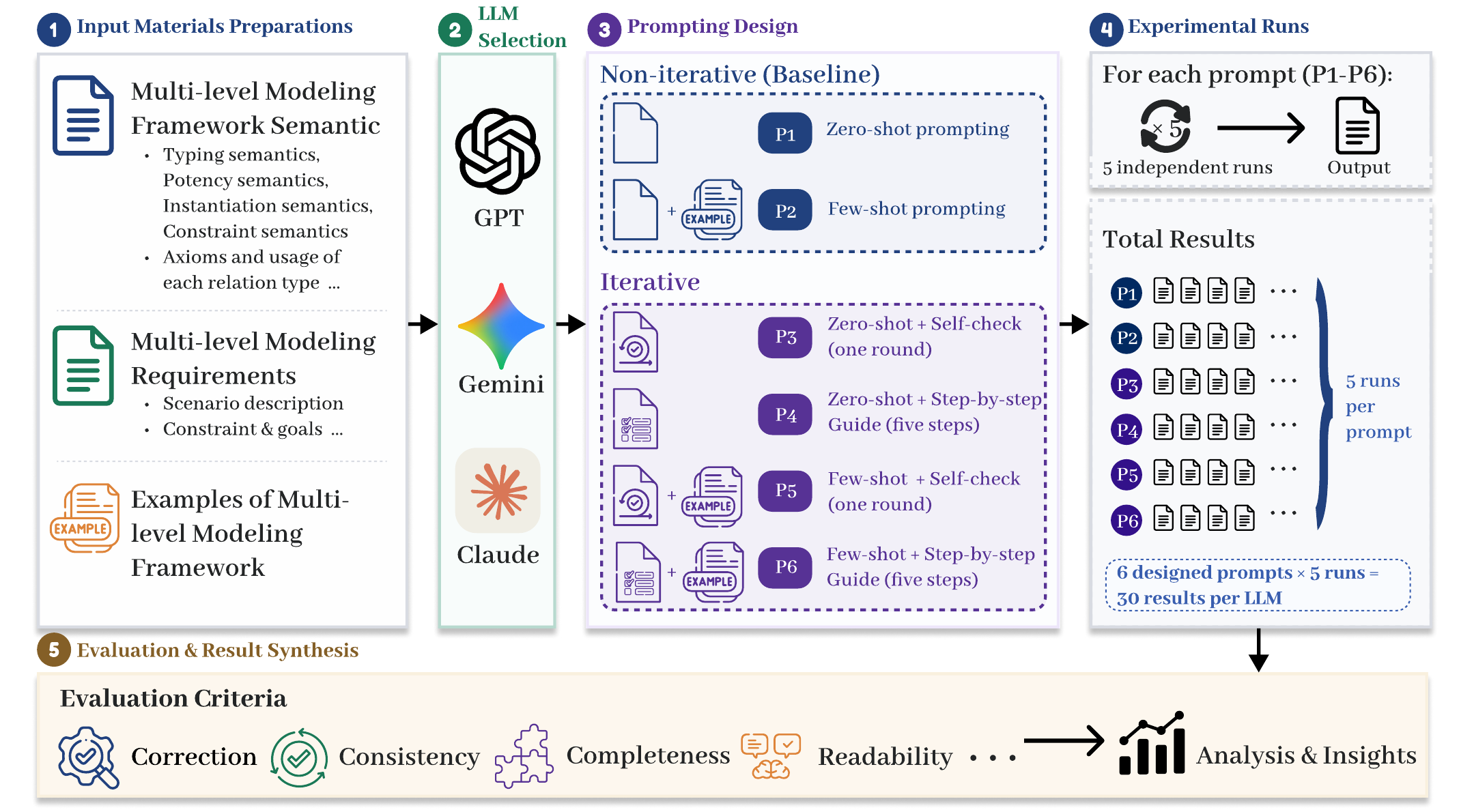}
\caption{Overview of the five-phase experimental design}
\label{fig:methodology}
\end{figure*}

This section describes the experimental design used to evaluate whether current LLMs can generate correct multi-level models in the SLICER language. The design follows the guidelines proposed by Baltes et al.~\cite{DBLP:journals/corr/abs-2508-15503} for empirical studies in software engineering involving LLMs. As shown in Figure~\ref{fig:methodology}, the study has five phases: input materials preparation, LLM selection, prompting design, experimental runs, and evaluation. These phases produce 90 generated models in total, obtained from three LLMs, six prompt configurations, and five repeated runs per configuration.


\subsection{Phase 1: Input Materials Preparation}
\label{sec:input}
All configurations use two core documents. \textbf{The first is a condensed modeling task description derived from the MULTI Warehouse Challenge~\cite{DBLP:conf/models/KuhneJ23}.} It describes the domain in natural language and specifies: (1) the product hierarchy from specification types through specifications to individual copies; (2) the attributes expected at each level, including which attributes are defined, constrained, or assigned values; (3) the relationships between elements, including which product specifications instantiate which specification types and which copies instantiate which specifications; (4) auxiliary structures such as product categories, warehouses, and storage locations; and (5) constraints that must hold across levels, including inventory traversal and price-related requirements.
\textbf{The second is a condensed SLICER reference derived from the SLICER journal article~\cite{DBLP:journals/dke/SelwaySMJGS17}}; it describes the seven SLICER relations, level derivation rules, attribute handling, constraint propagation, and relevant semantic axioms.

The few-shot conditions add one worked SLICER example from a different domain. This example shows how a complete SLICER model is expressed. All prompts, inputs, and generated outputs are included in the replication package~\cite{FuEtAl2026SoSyMSI}.

\subsection{Phase 2: LLM Selection}
We evaluate one frontier model from each of the three major commercial LLM families: GPT-5.4~\cite{openai2026gpt54} with reasoning effort set to \textit{extra high}, Claude Opus 4.6~\cite{anthropic2026opus46} with \textit{extended thinking} enabled, and Gemini 3.1 Pro~\cite{google2026gemini31pro} in its default configuration. These configurations were held fixed across all experimental runs reported in this paper. For readability, we therefore refer to the three models simply as GPT, Claude, and Gemini in the remainder of the paper. This choice gives broad provider coverage while keeping the experimental design manageable, a common trade-off in empirical studies of LLMs for software engineering~\cite{DBLP:journals/tosem/HouZLYWLLLGW24}. The models were selected because, at the time of experimentation, they were the strongest generally available representatives of their families and supported context windows large enough to hold the full SLICER semantic specification together with the task description in a single prompt.



\subsection{Phase 3: Prompting Design}
\label{sec:prompt_design}
Because SLICER is a low-resource modelling language, all prompts provide the language semantics in context. The prompts differ only in the input support and interaction structure given to the LLM, so that differences in output quality can be compared under controlled conditions.

\paragraph{Prompt Configurations}
We use a \(2\times3\) design: two input settings crossed with three interaction patterns. Table~\ref{tab:prompt-configurations} defines the six configurations and introduces the abbreviations used in the results.

\begin{table}[t]
\centering
\caption{Prompt configurations used in the experiment.}
\label{tab:prompt-configurations}
\small
\begin{tabular}{p{0.13\linewidth}p{0.34\linewidth}p{0.25\linewidth}p{0.18\linewidth}}
\toprule
\textbf{Abbrev.} & \textbf{Input setting} & \textbf{Interaction pattern} & \textbf{Evaluated output} \\
\midrule
ZS & Task + SLICER reference & Single-pass generation & First response \\
FS & Task + SLICER reference + example & Single-pass generation & First response \\
I-ZS-SC & Task + SLICER reference & Iterative self-check & Revised response \\
I-FS-SC & Task + SLICER reference + example & Iterative self-check & Revised response \\
I-ZS-SG & Task + SLICER reference & Iterative Step-by-step guide & Compiled model \\
I-FS-SG & Task + SLICER reference + example & Iterative step-by-step guide & Compiled model \\
\bottomrule
\end{tabular}
\end{table}

In the \textbf{single-pass} setting, the LLM receives the input materials and generates one final SLICER model as a markdown document.
In the \textbf{iterative self-check} setting, the interaction has three rounds: initial generation, checklist-based review, and revision. The checklist covers entity completeness, relation choice, level consistency, pricing and recommendation constraints, inventory traversal, and the terminal modelling of the bulk product. We keep both the draft and the revision, but only the revised model enters the main comparison; draft--revision differences are analysed separately in Section~\ref{sec:self-check-effects}.
In the \textbf{iterative step-by-step guide} setting, the interaction follows a five-step modelling workflow: identify domain entities, assign SLICER relations, define attributes and constraints, verify challenge-specific requirements, and compile the final model. Only the compiled model is evaluated. The same six configurations are used verbatim for all three LLMs, following the recommendation that LLM evaluations should use multiple prompt templates rather than relying on a single formulation~\cite{DBLP:journals/tacl/MizrahiKMDSS24}.

\subsection{Phase 4: Experimental Runs}

The three LLMs were accessed through a mixed setup: GPT through the OpenAI API, and Claude and Gemini through their provider-hosted web interfaces. Each LLM was run under each of the six prompt configurations, with five repeated runs per configuration, yielding 90 final models; for the two self-check configurations, the intermediate draft produced before revision was also retained, giving 120 generated models overall. For reproducibility, the replication package contains the reference model, the prompt materials for all six configurations (which embed the task description and the SLICER reference verbatim), all 120 generated models (the 90 final models and the 30 self-check drafts), the evaluation workbook with the per-model metric measurements, and the figure generation scripts together with the resulting PDF figures~\cite{FuEtAl2026SoSyMSI}. These artefacts provide traceability from each prompt configuration and repeated run to the corresponding generated model and reported evaluation results.

\subsection{Phase 5: Evaluation Metrics}\label{sec:metrics}

We compare each generated model \(M_g\) with a manually validated SLICER reference model \(M_b\). The metrics are grouped by their role in the analysis: validity metrics for RQ1, quality-comparison metrics for RQ2, and diagnostic evidence for the error-pattern analysis in RQ3. The complete per-model measurements are provided in the replication package.

Let \(O_g\) and \(O_b\) denote the generated and reference objects, \(R_g\) and \(R_b\) the generated and reference relations, \(A_b\) the reference attribute assignments, and \(G\) the union of generated objects and relations. Matching is semantic rather than literal: equivalent names, roles, and modelling intent count as matches, while a relation counts as correct only when both endpoints and the SLICER relation kind match. Unless stated otherwise, higher values are better; for error counts, edit distance, overgeneration, and redundancy, lower values are better. Repeated runs are reported as mean \(\pm\) standard deviation; differences between LLMs and prompting configurations are tested for statistical significance as described at the end of this subsection.

\paragraph{Syntactic correctness.}
For RQ1, Invalid Construct Count captures syntactic, typing, and SLICER well-formedness violations:
\[
\mathrm{ICC}(M_g)=\#\{\text{syntactic, typing, or SLICER well-formedness violations in }M_g\}.
\]
This includes malformed relation declarations, illegal relation usage, inconsistent level assignments induced by relation choices, and declared-but-unused constructs.

\paragraph{Semantic validity.}
RQ1 also requires semantic checks. Semantic Constraint Satisfaction Rate measures how many reference constraints \(C\) are represented in the generated model:
\[
\mathrm{SCSR}(M_g)=\frac{|C_{\mathrm{sat}}|}{|C|}.
\]
Instantiation/Specialisation Correctness focuses on the core SLICER distinction between InstN, InstX, SpecR, and SpecX:
\[
\mathrm{ISC}(M_g)=\frac{|S_{\mathrm{correct}}|}{|S_b|},
\]
where \(C_{\mathrm{sat}}\subseteq C\) is the subset of satisfied constraints, \(S_b\) is the set of reference instantiation and specialisation relations, and \(S_{\mathrm{correct}}\subseteq S_b\) is the subset reproduced with the correct relation kind.

\paragraph{Content coverage.}
For RQ2, we measure how much reference content each model and prompting strategy recovers. For objects and relations we report precision, recall, and F1:
\[
\mathrm{Precision}=\frac{|X_g\cap X_b|}{|X_g|},\quad
\mathrm{Recall}=\frac{|X_g\cap X_b|}{|X_b|},\quad
\mathrm{F1}=\frac{2\cdot\mathrm{Precision}\cdot\mathrm{Recall}}{\mathrm{Precision}+\mathrm{Recall}},
\]
with \(X\in\{O,R\}\). Attribute Accuracy measures the proportion of reference attribute assignments reproduced correctly:
\[
\mathrm{AttributeAccuracy}=\frac{|A_{\mathrm{correct}}|}{|A_b|}.
\]
Here \(A_{\mathrm{correct}}\subseteq A_b\) is the subset of reference attributes correctly reproduced by the generated model.
These metrics separate omission, overgeneration, and wrong relation-kind choices.

\paragraph{Structural similarity.}
We also use Graph Edit Distance and Hierarchy Preservation Rate to summarise global closeness to the reference design:
\[
\mathrm{GED}(M_g,M_b)=\#\{\text{node and edge insertions, deletions, and substitutions}\},
\quad
\mathrm{HPR}=\frac{|H_{\mathrm{correct}}|}{|H_b|}.
\]
Here \(H_b\) is the set of reference hierarchy links, and \(H_{\mathrm{correct}}\subseteq H_b\) is the subset preserved by the generated model. HPR is intentionally less strict than relation recall: it credits hierarchy edges with correct endpoints even when the SLICER relation kind is wrong. The gap between HPR and relation recall therefore indicates relation-kind confusion.

\paragraph{Extraneous structure.}
Finally, Hallucination/Overgeneration Rate and Redundancy Rate capture unsupported or duplicated modelling structure:
\[
\mathrm{HallucinationRate}=\frac{|G_{\mathrm{unsupported}}|}{|G|},\quad
\mathrm{RedundancyRate}=\frac{|G_{\mathrm{redundant}}|}{|G|}.
\]
Here \(G_{\mathrm{unsupported}}\subseteq G\) contains generated items not supported by the reference or task input, and \(G_{\mathrm{redundant}}\subseteq G\) contains duplicated or semantically overlapping items.
These RQ2 metrics also provide the evidence base for RQ3: the error-pattern analysis does not introduce a new score, but interprets the recurring semantic mistakes behind the validity and quality measurements.

\paragraph{Statistical analysis.} Because the per-cell samples are small (\(N=5\)) and several metrics are dominated by ties, we use non-parametric tests throughout, following established recommendations for randomised algorithms in software engineering~\cite{arcuri2014hitchhiker}: Kruskal--Wallis tests for the LLM and configuration main effects, aligned-rank-transform (ART) ANOVA~\cite{wobbrock2011art} for their interaction, and pairwise Mann--Whitney \(U\) tests with Holm--Bonferroni correction~\cite{holm1979simple}, reporting Cliff's \(\delta\)~\cite{cliff1993dominance} as effect size. Draft--revision effects of the self-check strategies are tested with Wilcoxon signed-rank tests on the 30 matched pairs. All tests are two-sided at \(\alpha=0.05\), and results are reported inline with the corresponding metrics. Redundancy Rate is not tested, as it has a single non-zero value among the 90 models.
Given the small per-cell sample size (N = 5), these tests are best read as a supplementary check against the possibility that observed patterns are pure chance, rather than as precise, standalone evidence of effect size or reliability; the descriptive statistics and figures reported throughout Section 4 remain the primary basis for our claims.

\section{Evaluation and Results}
\subsection{Benchmark Scenario and Reference Baseline}\label{sec:Evaluation_Setup}

This subsection presents the modelling scenario used to evaluate the LLMs, based on the MULTI Warehouse Challenge, a benchmark scenario designed by the MLM community to enable objective comparisons between different MLM approaches~\cite{DBLP:conf/models/KuhneJ23}.

\paragraph{Modelling Scenario}
The benchmark scenario is the MULTI Warehouse Challenge, which was designed by the MLM community to support comparisons between MLM approaches~\cite{DBLP:conf/models/KuhneJ23}. It describes a warehouse that sells two kinds of products:
\begin{itemize}
    \item copy-based products, namely books, DVDs, DVD players, mobile phones, and phone cases, whose physical copies are tracked individually; and
    \item bulk products, represented by AA battery cells, which are tracked by available quantity and pack size rather than by individual copies.
\end{itemize}

The challenge combines ordinary product modelling with multi-level constraints. Product specification types fix information such as tax rate, currency, and introduction date; product specifications fix standard sales prices; individual copies may add reduced prices and return dates. The required model must also support type-safe currencies, reduced prices below standard prices, derived final prices including tax, typed recommendation links, uniform inventory traversal over copies and bulk stock, and revenue sums per specification and per specification type~\cite{DBLP:conf/models/KuhneJ23}.

This scenario is suitable for evaluating MLM because some properties are fixed at one level but constrained or assigned at another. For example, a product specification type can require a price attribute without fixing all concrete price values. Capturing this constrained variability requires relation choices that preserve the intended level structure.
The LLM-facing version of this requirement is the condensed task document described in Subsection~\ref{sec:input}, which is included verbatim in the prompt materials of the replication package~\cite{FuEtAl2026SoSyMSI}; the complete challenge description is available in the original publication~\cite{DBLP:conf/models/KuhneJ23}.

\paragraph{Reference Model (baseline)}
The reference model is the published SLICER solution to the challenge~\cite{DBLP:conf/models/FuSGKS24}. Under the counting rules of Section~\ref{sec:metrics}, the evaluated reference baseline is:
\begin{itemize}
    \item \textbf{Objects:} 24 objects on four levels: the specification framework on the top level (ProductSpecType, ProductSpec, IndividualProductSpec, BulkProductSpec, MonetaryValue, Currency, and four currency instances); six concrete specification types on level 1; six product specifications on level 2; and two book copies on level 3.
    \item \textbf{Relations:} 27 SLICER relations: one SbS powertype link, eight SpecX relations, eleven InstX relations, and seven InstN relations.
    \item \textbf{Key modelling choices:} the bulk product is terminal and therefore uses InstN, whereas copy-based product specifications use InstX so that physical copies can be instantiated below them. Currency values are modelled as instances of Currency.
    \item \textbf{Evaluated constraints and attributes:} five constraints enter the evaluation: currency type safety, reduced price below standard price, equal currency for reduced and standard price, and non-negativity of both prices. The final-price rule is represented as a derived attribute and is evaluated with the attributes.
    \item \textbf{Exclusions:} general associations, including recommendation links, are modelled as attributes. Operations and their parameters are not compared. The reference solution does not use Cat or Member; generated category structures are therefore excluded from the scored comparison rather than counted as errors.
\end{itemize}

This setup focuses the evaluation on three modelling capabilities that are central to SLICER and to the Warehouse Challenge: choosing the correct relation kind, preserving the level structure induced by those relations, and representing properties that are defined at higher levels but assigned at lower levels.

\paragraph{Matching Rules}
Metrics are computed under the following matching rules. Element matching is semantic rather than literal: renamings of reference concepts (for example \textit{ProductSpecificationType} for \textit{ProductSpecType}, or \textit{Money} for \textit{MonetaryValue}) count as matches, and relation endpoints may be resolved through intermediate refinement classes introduced by the generated model. A relation counts as correct only if both endpoints and the SLICER relation kind match. Edges with correct endpoints but the wrong relation kind are treated as substitutions, counting towards Hierarchy Preservation but not towards Relation Precision or Recall. Since the reference solution does not use Cat or Member, generated category structures are excluded from all counts; operations and their parameters are likewise excluded from the attribute comparison.

\subsection{Results}\label{sec:Results}

This subsection reports the comparison of the 90 generated models against the reference model, using the prompt-configuration abbreviations defined in Table~\ref{tab:prompt-configurations}. All values are mean $\pm$ standard deviation over the \(N=5\) runs per configuration. For the two self-check strategies, the tables report the revised model produced after the check; the before/after comparison is given in Section~\ref{sec:self-check-effects}. 
Statistical test results (Kruskal–Wallis, Mann–Whitney U with Holm correction, Cliff's $\delta$) are reported alongside the descriptive statistics where available, but given the limited statistical power at this sample size, we treat them as corroborating detail rather than the primary basis for the claims made in this section.

\subsubsection{RQ1: Syntactic and Semantic Validity}\label{sec:results-rq1}

RQ1 asks whether LLMs can produce models that are syntactically well formed and semantically valid at both the SLICER-language and warehouse-domain levels. We therefore consider syntactic correctness, domain-level semantic correctness, and language-level MLM correctness.

\paragraph{Syntactic correctness.}
We assess syntactic correctness using \textit{Invalid Construct Count}. Lower values indicate better syntactic correctness.
\begin{figure}[t]
\centering
\includegraphics[width=0.9\linewidth]{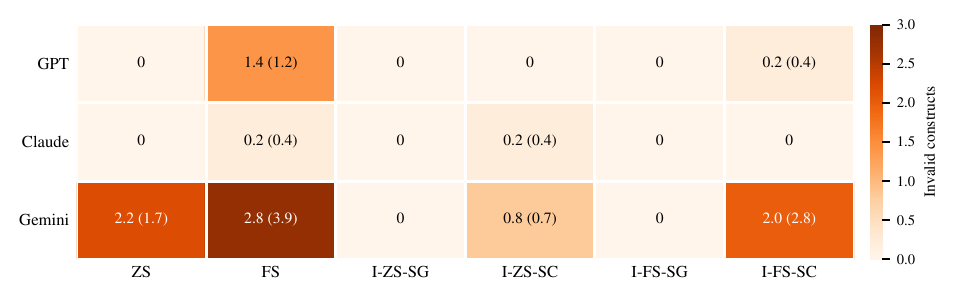}
\caption{Invalid Construct Count by LLM and prompting strategy. Cell values are means over the $N{=}5$ runs, with standard deviations in parentheses; darker cells contain more violations.  }
\label{fig:invalid-construct-count}
\end{figure}
All 90 final outputs could be interpreted as SLICER models, although some contained SLICER rule violations. With one exception, the models used only the seven SLICER relation names; the exception is one Claude self-check family that added a helper construct named \textit{Partitions} to express a covering condition, which we count as an invalid construct. Apart from this case, the violations in Figure~\ref{fig:invalid-construct-count} are breaches of SLICER rules rather than malformed syntax. The most common one is a SpecX relation between two elements on the same level, although SpecX must introduce a new level (axiom 37). It occurs mainly in Gemini outputs  (Gemini vs.\ Claude: \(\delta=0.32\), Holm-corrected \(p=0.010\)), which tend to declare copy/bulk branch classes via SpecX but keep them on level 0. The remaining violations are isolated: refinement relations (SpecR) that add attributes, constructs that are declared but never used, InstX steps that add no attributes, and one use of an undeclared attribute. Under both step-by-step variants, no model produced an invalid construct, and the self-check removed most violations present in the drafts (Section~\ref{sec:self-check-effects}).

\paragraph{Domain-level semantic correctness.}
We assess domain-level semantic correctness using the \textit{Semantic Constraint Satisfaction Rate}. Higher values indicate better semantic correctness.
\begin{figure}[t]
\centering
\includegraphics[width=0.75\linewidth]{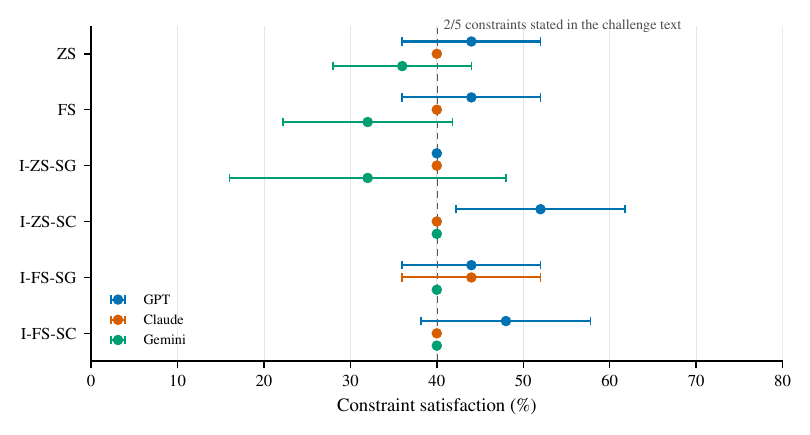}
\caption{Semantic Constraint Satisfaction Rate per configuration (mean $\pm$ SD over $N{=}5$ runs). The dashed line marks 40\%: two of the five reference constraints are stated in the challenge text.}
\label{fig:semantic-constraint-satisfaction}
\end{figure}
The satisfaction rate stays at or close to 40\% (Figure~\ref{fig:semantic-constraint-satisfaction}); it is flat across prompting strategies (Kruskal--Wallis, \(p=0.214\)), 
while small differences between LLMs are also present (Kruskal–Wallis \(p<0.001\); GPT vs. Gemini $\delta$=0.38, a modest effect).
The reason is visible in the raw data. Two of the five reference constraints are stated in the challenge text (currency type safety and the reduced price below the standard sales price), and nearly every generated model covers both, often together with a correct account of how the constraint propagates across levels. The other three are constraints the reference model adds on its own (non-negativity of the two prices, and the requirement that reduced price and standard price use the same currency), and the generated models almost never include them: across the 120 models available when the 30 self-check drafts are included, thirteen add exactly one of them, eleven of these GPT runs with a non-negativity check. Nine models, all from Gemini, cover only one of the five. No clear prompting-related pattern is visible for this metric: the models reproduce the constraints they are told about and rarely add protective ones. The final-price rule, which the challenge also states, is no longer part of this metric; it is represented in the reference as a derived attribute and is covered by the attribute comparison, where almost every model defines an equivalent derivation.

\paragraph{Language-level MLM correctness.}
We assess MLM correctness using \textit{Instantiation/Specialisation Correctness}. We do not evaluate level placement as a separate dimension: in SLICER, levels are derived from relations rather than assigned by the modeller, so level correctness follows from relation correctness, while violations of the level axioms are already captured by Invalid Construct Count above.
\begin{figure}[t]
\centering
\includegraphics[width=0.75\linewidth]{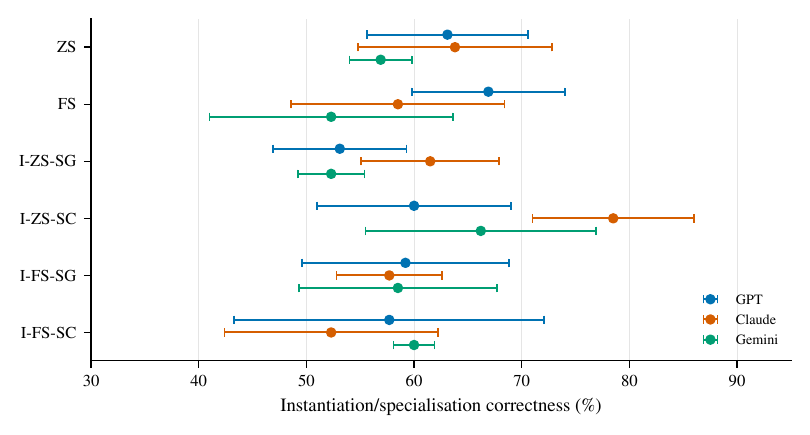}
\caption{Instantiation/Specialisation Correctness per configuration: share of the 26 baseline instantiation and specialisation relations reproduced with the correct relation kind (mean $\pm$ SD over $N{=}5$ runs).}
\label{fig:instantiation-specialisation}
\end{figure}
\textit{Instantiation/specialisation correctness} lies between 52\% and 79\% (Figure~\ref{fig:instantiation-specialisation}). The instantiation chain from specification types through product specifications to copies is reproduced with the correct kinds in nearly all models. The missing share comes mainly from specialisation edges into the branch classes, where SpecR and SpecX are frequently swapped, and from the four currency instantiations, which most models either omit or express as specialisations. A wrong relation kind also implies a wrong derived level structure: for example, connecting a branch class with SpecR instead of SpecX keeps it one level too high. No model or strategy resolves this: the three LLMs do not differ significantly (Kruskal--Wallis, \(p=0.419\)), and the strategy effect, though significant as an omnibus (\(p=0.040\)), leaves no pairwise contrast surviving Holm correction. Thus, the core domain chain is usually preserved, but correct application of SLICER's multi-level relation semantics remains partial.

\begin{rqanswer}{Answer to RQ1}
With the SLICER specification in context, syntactic correctness is within reach: all 90 final models could be interpreted as SLICER models, and rule violations were rare. Semantic validity is only partial. At the language level, Instantiation/Specialisation Correctness ranges from 52\% to 79\%, with errors concentrated in relation kinds that determine the multi-level structure. At the domain level, the models reliably enforce the two evaluated constraints stated in the task but almost never add the three defensive constraints supplied by the reference model. The generated models are therefore usable drafts, but they are not consistently complete or semantically reliable without expert checking.
\end{rqanswer}

\subsubsection{RQ2: Effects of LLM Choice and Prompting Strategy}\label{sec:results-rq2}

To answer RQ2, we compare the six prompting configurations across the validity metrics introduced in RQ1 and the quality dimensions reported below: content coverage, structural similarity, extraneous structure, self-checking behaviour, and run-to-run variability. The aim is to characterise the trade-offs associated with different LLM--prompt combinations.


\paragraph{Object-level completeness.}
We assess object-level results using \textit{Object Precision, Object Recall, and Object F1 Score}. These metrics evaluate whether the generated models contain the correct modelling objects.
\begin{figure}[t]
\centering
\includegraphics[width=0.62\linewidth]{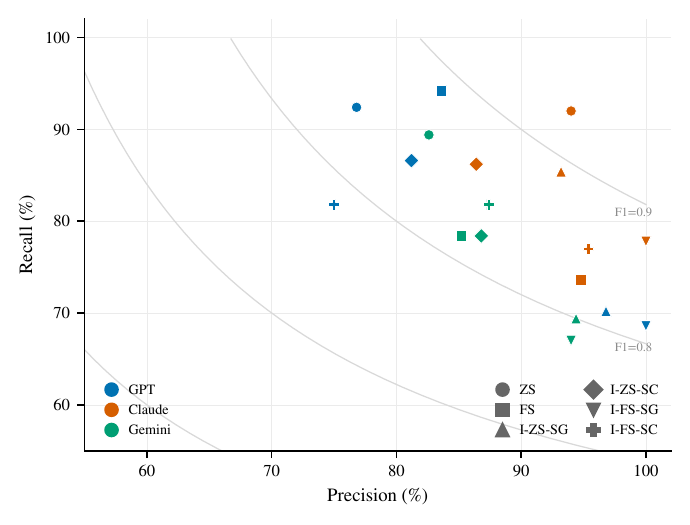}
\caption{Object precision vs.\ recall per configuration (means over $N{=}5$ runs). Grey curves are iso-F1 lines; colour encodes the LLM, marker shape the prompting strategy. The guided configurations (triangles) sit in the high-precision, low-recall corner, the unguided ones closer to the recall axis.}
\label{fig:object-precision-recall}
\end{figure}
Every one of the 120 generated models contains all 14 domain elements of the scenario: the six specification types, the six product specifications, and the two book copies. No model dropped or invented a product. Object-level differences therefore come entirely from the reference model's supporting structure: Currency and its four instances, MonetaryValue, and the two product-side branch classes. An equivalent of IndividualProductSpec appears in 38 of the 90 final models, of BulkProductSpec in 39, of MonetaryValue in 32, and of Currency in 38; the recall losses visible in Figure~\ref{fig:object-precision-recall} are concentrated there. Precision losses mirror this: extra objects are almost always additional scaffolding, most often currency-specific price types (\textit{EURPrice}, \textit{USDAmount}, and similar) and type-side copy/bulk classes with no counterpart in the reference. Figure~\ref{fig:object-precision-recall} shows the resulting trade-off: the step-by-step guides raise precision, up to 100\% for GPT and Claude under I-FS-SG, but lower recall to roughly two-thirds (both shifts significant: I-FS-SG vs.\ ZS, \(\delta=0.69\) for precision and \(\delta=-0.92\) for recall, Holm-corrected \(p\le0.011\)), because guided models build only the structure the guide walks through. The highest F1 is Claude under plain zero-shot (92.4\%); across configurations, F1 differs significantly by LLM (\(p=0.014\)) but not by strategy (\(p=0.394\)), as the guides trade precision against recall without a net F1 gain.

\paragraph{Relation-level completeness.}
We assess relation-level results using \textit{Relation Precision, Relation Recall, and Relation F1 Score}. These metrics evaluate whether the generated models capture the required edges with the correct relation kinds.
\begin{figure}[t]
\centering
\includegraphics[width=\linewidth]{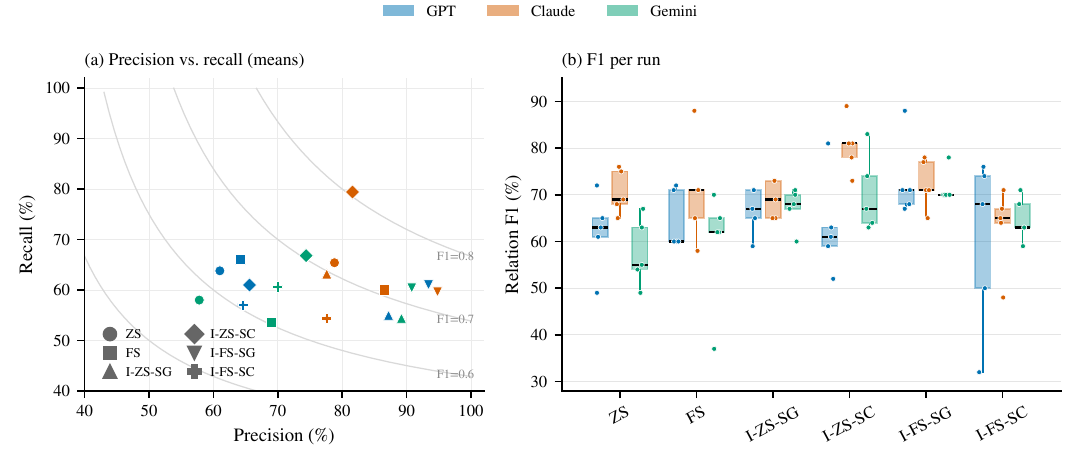}
\caption{Relation-level accuracy. (a)~Precision vs.\ recall per configuration (means over $N{=}5$ runs); grey curves are iso-F1 lines, marker shape encodes the strategy. (b)~Relation F1, one dot per run; boxes show median and quartiles. }
\label{fig:relation-accuracy}
\end{figure}

Relation scores are consistently lower than object scores, which confirms that selecting the right SLICER relation is harder than naming the right elements. Recall stays between 54\% and 79\%, with no significant effect of either LLM or strategy (\(p\ge0.112\)) (Figure~\ref{fig:relation-accuracy}(a)); the best cell is Claude with zero-shot self-check. Figure~\ref{fig:relation-accuracy}(b) shows the run-level spread behind these means. Two loss sources dominate. First, relations missing together with their objects: a model without currency instances cannot have the four InstN edges into Currency, and a model without branch classes cannot have the two SpecX edges into ProductSpec. Second, relations that connect the right elements with the wrong kind. Such substitutions do not count as correct here; the gap between Hierarchy Preservation and Relation Recall, which reaches 12 percentage points for the unguided strategies and nearly vanishes under the guides, shows how common they are. The most frequent substitutions are SpecR for SpecX (or the reverse) in the dual classification of the six specification types, and SpecR or SpecX where the reference instantiates the four currencies with InstN. The same precision--recall trade-off appears at relation level: under I-FS-SG all three LLMs emit almost only correct edges (91--95\% precision; vs.\ ZS \(\delta=0.92\), \(p<0.001\)) while reproducing just over half of the reference relations.

\paragraph{Attribute-level coverage.}
We assess attribute-level coverage using \textit{Attribute Accuracy}. This metric evaluates whether matched model elements preserve the correct attribute values, labels, properties, or metadata.


\begin{figure}[t]
\centering
\includegraphics[width=0.8\linewidth]{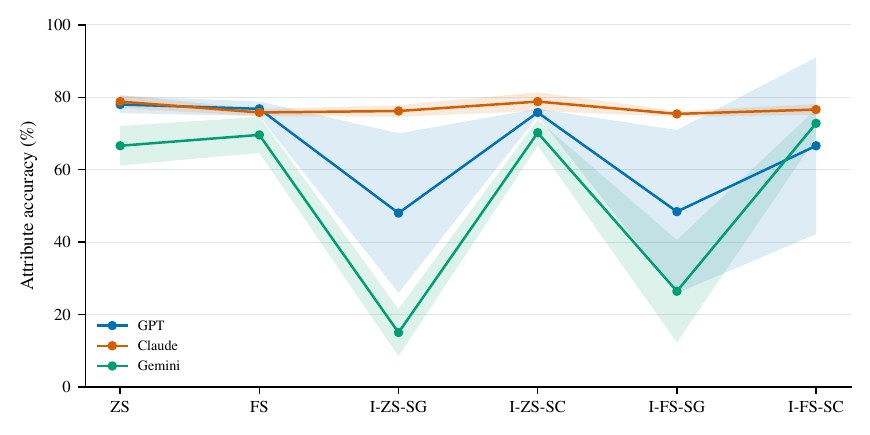}
\caption{Attribute accuracy across prompting strategies (lines connect means; bands are $\pm$ SD). }
\label{fig:attribute-accuracy}
\end{figure}

For the strategies that produce complete models, attribute accuracy lies between 67\% and 79\% (Figure~\ref{fig:attribute-accuracy}). Part of the loss is shared by all models and traces back to attributes that appear only in the reference figure, not in the challenge text: \textit{accessoriesMissing} and \textit{bookCat} occur in none of the 90 models, the \textit{itemsSold} counter in six (never with its reference value of 500), and the \textit{sold} flag in nine. Beyond that, scenario values (prices, tax rates, currencies, dates, stock quantity) are transcribed correctly whenever a model prints them at all; we found no wrong value in any model. The low scores of GPT and Gemini under the step-by-step guides (15--48\%) have a specific cause: their guided outputs name the attribute schema but assign concrete catalogue values only for the worked example, the Moby Dick chain, and leave the rest of the catalogue unstated. Claude is the only model that keeps printing the full value set under every strategy, which is why its attribute accuracy is stable across columns; this model dependence yields the strongest LLM~($\times$)~strategy interaction of any metric (ART ($F_{10,72}=10.33$), \(p<0.001\)).

\paragraph{Structural similarity.}
We assess structural similarity using \textit{Graph Edit Distance} and \textit{Hierarchy Preservation Rate} between generated and reference models.


\begin{figure}[t]
\centering
\includegraphics[width=0.62\linewidth]{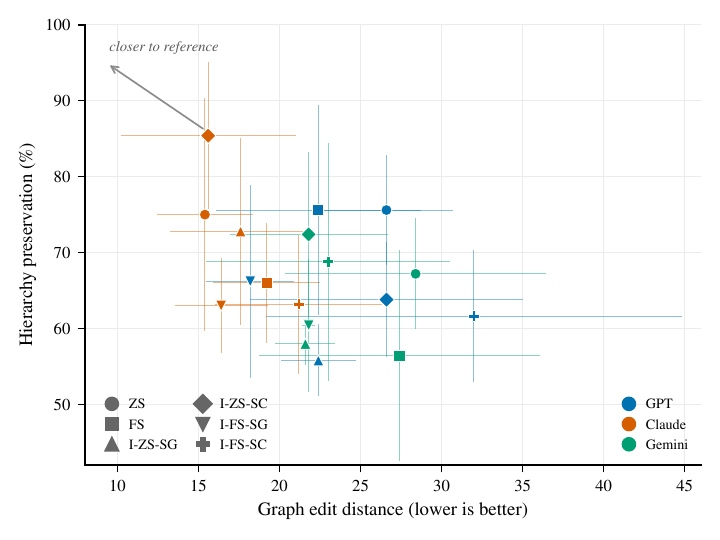}
\caption{Graph edit distance vs.\ hierarchy preservation per configuration (means over $N{=}5$ runs; error bars $\pm$ SD); colour encodes the LLM, marker shape the strategy. Points towards the upper left are closer to the reference model.}
\label{fig:structural-similarity}
\end{figure}

Against a baseline of 24 objects and 27 relations, the mean distance to the reference ranges from about 15 to 32 edit operations (Figure~\ref{fig:structural-similarity}). Claude needs the fewest edits in five of the six strategies, with the overall best value (15.4) under plain zero-shot; the advantage is significant (vs.\ GPT \(\delta=-0.59\), vs.\ Gemini \(\delta=-0.58\), both \(p<0.001\)), while the strategy has no significant effect on distance (\(p=0.322\)). GPT's distance rises under few-shot self-check because one revision added eleven kind-specific classes, and Gemini's zero-shot distance reflects its habit of adding a parallel product-copy class hierarchy. Hierarchy preservation exceeds relation recall everywhere, since it also credits edges drawn with the wrong relation kind. The type--instance chain from specification types through specifications to copies is preserved in nearly all models; what is lost concerns the supporting structure, not the domain hierarchy.

\paragraph{Overgeneration and redundancy.}
We assess extraneous structure using \textit{Hallucination/Overgeneration Rate} and \textit{Redundancy Rate}. Lower values indicate better performance.



\begin{figure}[t]
\centering
\includegraphics[width=\linewidth]{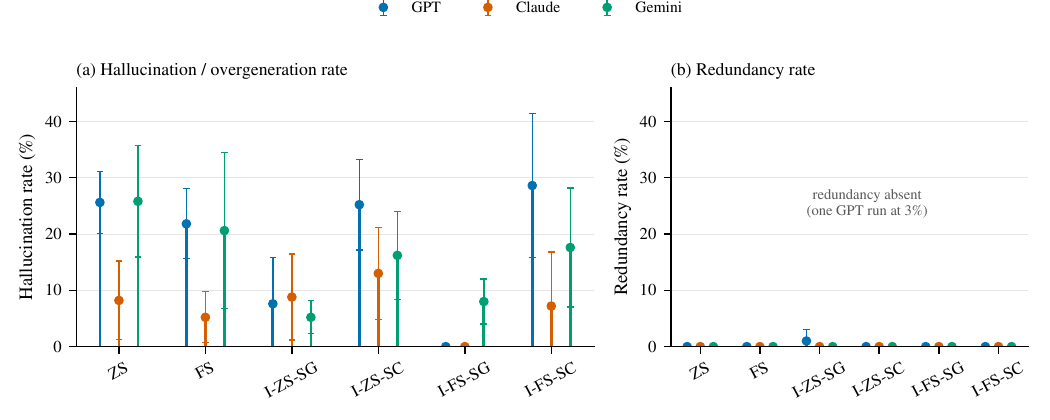}
\caption{(a)~Hallucination/overgeneration rate and (b)~redundancy rate per configuration (means; whiskers $\pm$ SD).}
\label{fig:overgeneration-redundancy}
\end{figure}

Overgeneration is moderate and almost entirely structural (Figure~\ref{fig:overgeneration-redundancy}). With two exceptions across the 90 final models, one hypothetical extra product copy and one assumed external sales ledger, every unsupported element is scaffolding: currency-specific money types, type-side copy/bulk specification-type classes, explicit product-copy classes, or sales and portfolio bookkeeping entities. Claude adds the least (0--13\%; vs.\ GPT \(\delta=-0.51\), \(p=0.002\); vs.\ Gemini \(\delta=-0.45\), \(p=0.005\)), GPT and Gemini add more in unguided settings (about 20--30\%), and the step-by-step guides suppress the behaviour almost completely (I-FS-SG vs.\ ZS, ($\delta=-0.79$), (p=0.002)). Redundancy is essentially absent (Figure~\ref{fig:overgeneration-redundancy}(b)); the single case in 90 models is one GPT run that kept two objects in the product-specification role.

\paragraph{Effect of self-checking and run-to-run consistency.}\label{sec:self-check-effects}
Comparing the 30 draft/revision pairs, the check works primarily as a rule checker. Mean invalid construct counts fall from up to 5.6 per draft (Gemini, few-shot) to at most 2.0 after revision, all five wrong bulk-product instantiations present in the drafts are fixed, and the only violation that survives a check in Claude's runs is the invented \textit{Partitions} construct, which the check does not question. The relationship with similarity to the reference is small and mixed: relation F1 moves by a few points in either direction, and several revisions make the model larger by adding kind-specific classes, so \textit{Graph Edit Distance} sometimes becomes worse. One round of self-checking, therefore, reduces rule-level violations, but it does not reliably move a model closer to the reference design. A Wilcoxon signed-rank test on the 30 pairs supports this reading: the reduction in invalid constructs is significant (\(p<0.001\), rank-biserial \(r=-0.93\)), while the paired changes in relation F1 (\(p=0.094\)) and Graph Edit Distance (\(p=0.328\)) are not.

Run-to-run variability is reflected in the reported standard deviations and run-level distributions. No prompting configuration is uniformly the most stable across all LLMs and metrics. Guided configurations have better rule conformance in this benchmark, but their relationship with completeness, attribute correctness, and structural similarity remains model-dependent.

\paragraph{Cross-metric comparison.}
All 90 final models contain the 14 domain elements in the scenario, so the three LLMs do not differ in basic domain coverage; their differences emerge in the supporting structure, choice of SLICER relation kinds, amount of detail, and stability. Claude has the most balanced profile: it reaches the highest object F1 (92.4\% under zero-shot prompting), has the lowest Graph Edit Distance in five of the six prompting configurations (with the overall minimum of 15.4 under zero-shot prompting), keeps attribute accuracy between 75\% and 79\% under every strategy, and produces almost no rule violations. GPT remains competitive on relation quality but introduces the most structural scaffolding and varies most across runs and prompting strategies; guided output also omits many catalogue values. Gemini accounts for most same-level SpecX violations, prints the fewest attribute values, and has the lowest constraint coverage, although guidance markedly improves its precision and suppresses overgeneration. 
These profile differences are consistent with the descriptive picture above and hold up under testing: Claude's leads on object F1 (vs.\ Gemini, $\delta = 0.41$), relation F1 (vs.\ both, $\delta = 0.37$--$0.42$), edit distance, and over generation all survive Holm correction ($p \leq 0.030$), and Gemini's attribute gap is the largest effect observed ($\delta \geq 0.61$ against both, $p < 0.001$), though these effect sizes should be read in light of the sample size noted in Section~\ref{sec:threats}.

\begin{rqanswer}{Answer to RQ2}
Prompting strategy and LLM choice both have significant effects on most quality dimensions, but no strategy dominates across all metrics. Step-by-step guidance raises precision (up to 100\%) and removes almost all invalid constructs and overgeneration, but lowers recall (to roughly two thirds) and, for GPT and Gemini, catalogue completeness; few-shot examples improve structure but introduce a systematic InstX error absent from zero-shot output; self-checking fixes rule violations without moving models closer to the reference design. Across LLMs, Claude has the most balanced profile in this benchmark, GPT overgenerates the most scaffolding, and Gemini shows the most rule violations and sparsest attribute coverage. The observed gains from guidance are largest for Gemini on precision and rule conformance, whereas Claude changes little across strategies. This suggests that prompting can partly compensate for weaker baseline performance in this setting; the significant LLM~($\times$)~strategy interactions (strongest for attribute accuracy) support this model dependence, though confirmation on a broader set of tasks and models remains necessary.
\end{rqanswer}


\subsubsection{RQ3: Common Semantic Error Patterns}\label{sec:error-patterns}

Across the 90 final models, \textbf{six recurring semantic error patterns} explain most of the quality loss. Where self-check drafts are relevant, we additionally report the corresponding draft counts. Table~\ref{tab:error-patterns} summarises the patterns; the discussion below groups them by their underlying cause.

\begin{table}[t]
\centering
\caption{Recurring semantic error patterns in the generated SLICER models.}
\label{tab:error-patterns}
\footnotesize
\setlength{\tabcolsep}{4pt}
\renewcommand{\arraystretch}{1.16}
\begin{tabular}{@{}>{\raggedright\arraybackslash}p{0.27\linewidth}>{\raggedright\arraybackslash}p{0.67\linewidth}@{}}
\toprule
\textbf{Error pattern} & \textbf{Evidence and metric implication} \\
\midrule
Terminal elements made extensible
& \textit{Construct:} InstX used where the reference baseline uses InstN.\newline
  \textit{Evidence:} 10/90 final models: 8 bulk-product cases and 2 returned-copy cases.\newline
  \textit{Effect:} wrong terminality; lower ISC. \\
\addlinespace[2pt]
Currencies not modelled as instances
& \textit{Construct:} missing InstN links from Currency to EUR/USD/SEK/NZD.\newline
  \textit{Evidence:} 13/90 reproduce all four links; 17/90 model currencies as specialisations.\newline
  \textit{Effect:} missing objects and relations; lower recall. \\
\addlinespace[2pt]
Product-side copy/bulk split omitted
& \textit{Construct:} missing IndividualProductSpec/BulkProductSpec via SpecX.\newline
  \textit{Evidence:} branch objects appear in 38--39/90; full dual classification in 10/90.\newline
  \textit{Effect:} missing supporting structure. \\
\addlinespace[2pt]
Same-level SpecX
& \textit{Construct:} SpecX used without introducing a new level.\newline
  \textit{Evidence:} main invalid-construct category; 17/90 final models contain at least one invalid construct.\newline
  \textit{Effect:} lower well-formedness. \\
\addlinespace[2pt]
SpecR/SpecX substitution
& \textit{Construct:} wrong specialisation kind with otherwise relevant endpoints.\newline
  \textit{Evidence:} visible in the gap between HPR and Relation Recall.\newline
  \textit{Effect:} hierarchy partly preserved, but relation kind incorrect. \\
\addlinespace[2pt]
Unstated defensive constraints omitted
& \textit{Construct:} reference-only price constraints absent.\newline
  \textit{Evidence:} 10/90 final models, or 13/120 including drafts, add exactly one.\newline
  \textit{Effect:} lower SCSR. \\
\bottomrule
\end{tabular}
\end{table}

The common thread is that the models usually recover the stated product chain but struggle with SLICER-specific semantics that are only implicit in ordinary domain wording. Terminal objects are sometimes left extensible, currencies are often treated as datatypes or specialisations rather than instances, and the copy/bulk distinction is frequently placed on the type side rather than on the ProductSpec side used by the reference baseline. These choices create missing supporting objects and relations, and they also explain why hierarchy preservation can be higher than relation recall: several outputs connect the right conceptual elements but choose the wrong relation kind.
The remaining two patterns concern rule conformance and implicit constraints. Same-level SpecX is the clearest well-formedness violation because SpecX must introduce a new level; it is also the violation most directly addressed by self-checking. The omitted defensive constraints have a different character: the generated models usually reproduce the two evaluated constraints stated in the challenge, but rarely add the three constraints supplied only by the reference baseline. Thus, the errors are not failures to identify the visible product domain; they are failures to infer the supporting multi-level structure and unstated semantic safeguards.

\begin{rqanswer}{Answer to RQ3}
The recurring errors concern multi-level semantics and implied supporting structure rather than the stated product content. The models commonly preserve the visible specification-to-copy chain, but they often choose the wrong terminality, omit or remodel currency instances, miss the product-side copy/bulk split, confuse SpecR and SpecX, or omit defensive constraints not stated explicitly in the task. These patterns explain why the generated models can look domain-complete while still falling short of the SLICER reference baseline.
\end{rqanswer}

\section{Discussion}\label{sec:Discussion}
The results draw a fairly clear line between what current LLMs can reliably take on and what still requires human oversight. The sections that follow examine why this line falls where it does (\ref{sec:unified_account}, \ref{sec:why_rel_kind}), situate it against existing evidence from two-level modelling (\ref{sec:2lm_vs_mlm}), and discuss its implications for practical choices and for the vision of human-in-the-loop modelling (\ref{sec:practical_implications}, \ref{sec:implications_4_human_certric}).

\subsection{A Unifying Account: Explicit versus Implicit Structure}
\label{sec:unified_account}
The three findings reported above appear separate, but they point to the same underlying regularity: the generated models faithfully reproduce what the task text states explicitly, while rarely completing the structure that the task implies but does not spell out, and that a modeller would normally need to infer.

The first piece of evidence concerns constraints. The two constraints stated explicitly in the challenge text (currency type safety, and the reduced price staying below the standard sales price) are covered by nearly every generated model, while the three defensive constraints that the reference model adds on its own (non-negativity of both prices, and matching currencies between the reduced and standard price) are almost never included. This is not a matter of models omitting a random subset of constraints; what is missing is precisely the category of constraint that the text does not state outright.
The second piece of evidence concerns structure. The reference model splits ProductSpec into IndividualProductSpec and BulkProductSpec. This split is not given directly as a class in the text; it must be inferred from the description of copy based and bulk products. More than half of the generated models fail to make this split, and instead add alternative scaffolding on the type side.
The third piece of evidence is also structural. The reference model instantiates the four currencies from Currency using InstN. This design is likewise not stated directly in the text; it follows from working backward from the currency type safety constraint. Most generated models either treat currency as a plain attribute value or express it through specialisation rather than instantiation, so currency as a first class object, together with its instantiation relations, is largely absent.

Together, these three observations suggest that current LLMs act in this task more as faithful transcribers than as modellers who complete gaps in a specification on their own initiative. Content that is written into the text, whether constraints, entities, or relations, is reliably converted into SLICER constructs, while structure that is not written into the text, and that requires modelling experience or backward reasoning from a stated constraint to supply, is systematically omitted or replaced with scaffolding of the model's own design. This distinction accounts for the pattern of failures observed across RQ1 to RQ3 more directly than a list of individual metric scores, and it provides a shared starting point for the discussion of prompting strategy and relation semantics that follows in the next two sections. Because this distinction concerns how an LLM completes a specification in general, rather than any feature specific to SLICER, it offers one lens through which the results of this study speak to LLM-assisted modelling beyond the multi-level setting examined here.

\subsection{Why Relation-Kind Semantics Resist In-Context Learning}
\label{sec:why_rel_kind}
If 5.1 shows that implicit structure is easily omitted, the errors in relation kind selection reveal a different kind of difficulty: even when elements are correctly identified and both endpoints of a relation are correctly connected, the relation kind chosen by the LLM is still often wrong. These errors concentrate between SpecR and SpecX, and between InstN and InstX, which are exactly the semantic extensions that distinguish SLICER from two-level modelling. This matches the comprehension-generation gap noted in Section 2.1.1: Liu et al.~\cite{DBLP:conf/acl/LiuCSZNH0L24} found that LLMs generally understand the structure of formal languages without difficulty, but perform considerably worse at generation. The results of RQ1 and RQ3 are consistent with this pattern: the core instantiation chain (from specification type through specification to copy) is reproduced correctly in nearly all models, showing that LLMs do grasp the overall type-instance direction that SLICER relations express. Once a precise choice must be made among relation kinds with closely related semantics, however, accuracy drops noticeably, with Instantiation/Specialisation Correctness falling only between 52\% and 79\%.

One possible explanation is that, at generation time, LLMs tend to collapse the fine-grained relation kinds of SLICER back into the single, more common pair of binary relations found in mainstream training corpora, namely inheritance and instantiation as used in standard object-oriented modelling. The distinction between SpecR and SpecX turns on whether a relation crosses an abstraction level and extends the attribute set; the distinction between InstN and InstX turns on whether an instance can be further instantiated. Neither distinction has a direct counterpart in two-level modelling, because both are semantic refinements that SLICER introduces specifically to support deep instantiation, where a single element simultaneously acts as a type and as an instance. We conjecture that this dual role is comparatively rare in LLM pretraining corpora, since the dominant languages for two-level modelling, such as UML, do not generally require an element to hold both roles at once. If this is the case, it may help explain a specific pattern observed in RQ3: SpecX being used between two elements at the same level, in violation of the axiom that SpecX must introduce a new level, a pattern consistent with fine-grained, multi-valued semantics being compressed back into a coarser and more familiar binary opposition.

The treatment of currencies offers additional support for this hypothesis. The reference model instantiates the four currencies from Currency using InstN, a terminal relation that does not permit further instantiation, while most generated models instead express this relationship through specialisation. Specialisation is the more common and, in a sense, the safer default choice in two-level modelling, which is consistent with the broader hypothesis that when the choice of relation kind is uncertain, LLMs fall back on the most frequent relation type in the corpus rather than deriving the semantically correct kind from SLICER's axioms.
This explanation remains a hypothesis grounded in observed patterns rather than a directly verified mechanism; the design of this study cannot confirm whether such a collapsing process actually occurs inside the model, nor did it examine the composition of the training data, and both are left as questions for future work. It nonetheless offers a coherent perspective on why relation-kind selection is the hardest part of the task in this study, and it leaves an open direction for Section 5.6, where we consider whether other MLM mechanisms, such as potency-based approaches, might face a similar difficulty.

\subsection{Two-Level versus Multi-Level Modelling Compared}
\label{sec:2lm_vs_mlm}
Locating the failure patterns of the SLICER task within the coordinates of MLM calls for a natural point of reference: the quantitative baseline already established in two-level modelling research. Chen et al.~\cite{chen2023automated} compared GPT-3.5 and GPT-4 on domain modeling, a task that is essentially class diagram generation in two-level modelling, and found that the best-performing LLM achieved F1 scores of 0.76 for classes, 0.61 for attributes, and only 0.34 for relationships, with precision generally exceeding recall, meaning that the generated elements were themselves reliable but many were missing. 
A broader comparison across thirteen LLMs and five prompting strategies by Calamo et al.~\cite{calamo2025assessing} reports the same asymmetry at a larger scale: the best class-detection F1 reached 0.660, while the best association F1 reached only 0.317, a gap the authors attribute partly to class extraction being reducible to a form of named entity recognition, whereas association and cardinality extraction have less direct precedent in the data LLMs are pretrained on.
This pattern aligns closely with the results of the present study: all 90 models correctly covered the 14 domain elements, the highest object-level F1 reached 92.4\%, yet relation-level scores were consistently lower (recall ranging from 54\% to 79\%), showing the same structure in which precision exceeds recall and relations are harder to recover than objects.

This alignment across two independent studies, different models, datasets, and prompting strategies, points to a pattern consistent with our observations: regardless of how complex the abstraction levels of a modelling language become, an LLM's ability to recover entities themselves appears to consistently exceed its ability to recover the relations between them, and correctness at the relation level may be a more general bottleneck than the identification of domain elements.

MLM, however, adds an additional category of error on top of this shared bottleneck, one that does not exist in two-level modelling at all. In the evaluation scheme of Chen et al., relationships fall into only three types (associate, inherit, contain), and errors mainly take the form of a missing relation or a wrong judgment of whether one exists at all. The relation-kind confusion observed in RQ3 of this study (the mistaken choice between SpecR and SpecX, or between InstN and InstX) occurs even when the relation itself is correctly identified and both endpoints are correctly connected, yet the wrong, more finely distinguished relation kind is still selected. This type of error has no counterpart in the evaluation framework of two-level modelling, because two-level modelling does not require a fine-grained choice among multiple relation types with closely related semantics.

A more accurate characterisation, then, is not that MLM makes LLMs perform worse overall, but that MLM adds a layer of relation-kind selection errors, absent from two-level modelling, on top of a relation-recovery bottleneck already known from two-level modelling. 
This comparison rests on a qualitative analogy across different tasks, models, and evaluation schemes, and does not support a direct statistical comparison; a definitive answer awaits future work that evaluates two-level and multi-level tasks side by side under the same experimental design.

\subsection{Practical Implications: Choosing a Prompting Strategy}
\label{sec:practical_implications}
The results of RQ2 can be translated into a more practice-oriented question: given the goal of obtaining a usable SLICER draft, which prompting strategy should be chosen.

The results point to two different paths, corresponding to two different purposes of use. If the goal is to produce a draft for a human expert to complete and correct, zero-shot or few-shot prompting is more suitable: both achieve higher recall, and since the missing parts require human intervention regardless, the shortfall in precision is not the main cost. If the goal is to obtain output that is rule-compliant and can feed directly into further automated processing, step-by-step guidance is more suitable: this strategy produces the lowest rate of invalid constructs, but as shown in RQ2, this comes at the cost of recall dropping to roughly two thirds, along with a substantial drop in attribute completeness for
GPT and Gemini. This trade-off should be understood before adopting the strategy.

Strategy selection also involves a dimension not directly quantified in RQ2: interaction cost. Zero-shot and few-shot require only a single round of interaction, self-check requires three rounds, and step-by-step guidance requires five. Yet the improvement brought by self-check is limited to fixing rule violations, with little effect, and sometimes a negative one, on overall similarity to the reference design; and while step-by-step guidance brings a clear gain in precision, it brings an equally clear loss in completeness. Whether the additional rounds of interaction are worth the investment depends on whether the resulting gain in precision is actually the benefit the task at hand needs, rather than assuming that more rounds automatically means higher quality.

This study therefore offers not a single optimal strategy, but a set of grounds for choosing one according to purpose: draft-oriented tasks should favour zero-shot or few-shot prompting, compliance-oriented tasks are the ones worth the additional cost of step-by-step guidance, and self-check is better suited as a supplementary rule-checking device rather than a means of improving structural similarity to the reference design.

\subsection{Implications for Human-Centric Modelling and Industry 5.0}
\label{sec:implications_4_human_certric}
As noted in the Introduction, MLM already embodies the human-centric design goal of Industry 5.0~\cite{breque2021industry5}: it shifts the burden created by the accidental complexity of two-level modelling~\cite{DBLP:journals/sosym/AtkinsonK08} away from the engineer and onto the modelling language itself~\cite{DBLP:conf/uml/AtkinsonK01}. The question posed at the outset of this paper is whether LLMs can carry this shift one step further, by taking on the correct understanding and application of semantic rules as well.

The results show that this shift has only been realised halfway. The core instantiation chain, that is, how the level structure should be built, is reproduced correctly in nearly all models, showing that LLMs do take on this part of the burden. But the precise choice of relation kind, the rule-based knowledge needed to keep a model semantically correct, still fails frequently (Instantiation/Specialisation Correctness of only 52\% to 79\%), and RQ1 states explicitly that the generated models are "not consistently complete or semantically reliable without expert checking." The burden of semantic correctness has not disappeared; it has only moved. It is no longer the engineer who must master SLICER's axioms while modelling, but the engineer, or some automated checking tool, who must master the same axioms afterward to review what the LLM has produced. The burden has been shifted, not absorbed.

This result need not be read as a refutation of the Industry 5.0 vision. Human-centric design was never meant to equate to full automation; it emphasises reducing cognitive burden while keeping humans involved in oversight at critical decision points~\cite{breque2021industry5}. Seen in this light, the reliable performance of LLMs on level structure, set against their unreliable performance on relation-kind selection, marks out a boundary for the division of labour between human and machine: structural burdens that a tool can absorb outright may reasonably be handed to an LLM, while decisions that depend on fine-grained semantic judgement, and whose errors are hard to detect, still require the involvement of a human expert or an automated checking tool. Recognising this boundary, rather than simply pursuing a complete transfer of burden, may be the more realistic position for LLMs in human-in-the-loop modelling, and one closer to what Industry 5.0 originally intended.

\subsection{Threats to Validity}
The following discusses the limitations of this study organised by the four standard categories of construct, internal, external, and conclusion validity.

\label{sec:threats}
\paragraph{Construct Validity}
The main threat is that our metrics compare each generated model against a single reference solution, and some deviations counted as errors by the metrics are in fact defensible design alternatives, for example modelling currencies as refinement subtypes of a price type instead of as instances, or placing the copy/bulk split on the type side; such alternatives lower precision and recall without being wrong in an absolute sense. It should be noted that the uniqueness of the reference model argued for elsewhere in this paper concerns only the portion of the model directly constrained by SLICER's axioms (for example, the level at which an attribute resides once it is assigned a value, which the axioms determine uniquely), while design decisions that the axioms do not force still admit such alternatives. We reduce this threat by matching elements semantically rather than by name, by resolving relation endpoints to intermediate refinement classes, and by excluding Cat/Member structures that the reference does not use; these measures narrow, but do not fully remove, the bias introduced by anchoring the metrics to a single solution.

A further threat is that the Semantic Constraint Satisfaction Rate depends on the constraints declared by the reference, and three of its five constraints are not stated in the challenge text. This means the metric is capped at 40\% for any model that follows the text alone. We did not adjust this metric, because this ceiling serves a diagnostic purpose: it separates models that merely reproduce what the text states from models that supply defensive constraints the text does not state, which is one of the central findings discussed in Section~\ref{sec:results-rq1} and Section~\ref{sec:unified_account}. The absolute value of this metric should therefore be read in light of that distinction, rather than treated on its own as an overall measure of model quality.

A last threat is that extracting objects, relations, and attributes from the generated documents required manual normalisation, which may introduce extraction errors. We reduce this risk through two rounds of verification; the per-model metric measurements underlying all reported results are included in the replication package.

\paragraph{Internal Validity}
Existing threats include the non-determinism of LLM outputs (mitigated by running each configuration N=5 times) and the possibility that prompt design favors certain LLMs over others (mitigated by using identical prompts across all models).

The scoring process itself also constitutes a threat. Because element matching is semantic rather than based on string comparison, extracting objects, relations, and attributes from the generated documents required manual judgement (as described in Section~\ref{sec:Results}). The scorers were not blind to the source of a given model output (GPT, Claude, or Gemini) while evaluating it, which could introduce rater bias, particularly since this paper reports a relative ranking across models. This risk is reduced by the fact that scoring follows the quantitative rules predefined in Section~\ref{sec:prompt_design}, for example whether a relation kind matches the reference, which are comparatively objective judgements with limited room for subjective discretion, but the absence of blind scoring remains a limitation of this study.

In addition, the SLICER semantic reference document, the Warehouse requirements document, and the reference model used for comparison were all authored by, or previously published by, the authors of this paper (the reference model corresponds to the solution described in~\cite{DBLP:conf/models/FuSGKS24}). The authors who designed the prompt materials are also the authors of the reference model, which may introduce an unconscious bias in judging what counts as implicit structure that the LLM must infer on its own. The scope of this influence is limited: the requirements text and SLICER specification content are given verbatim in the prompts, which are included in the replication package, and are therefore fixed and auditable, so this threat mainly concerns the evaluation design rather than uncontrolled variation in the content of the prompts themselves.

\paragraph{External Validity}
The main threat is that this study uses a single modelling task (the Warehouse Challenge), and the results may not generalise to other domains or other modelling tasks. We mitigate this risk by choosing the MULTI Warehouse Challenge, a benchmark scenario designed by the MLM community specifically to enable objective comparisons between different MLM approaches (as described in Section~\ref{sec:Evaluation_Setup}), rather than a task constructed for this study alone.
This study also involves a single MLM language (SLICER), and its conclusions are therefore limited in scope to the relation-driven class of MLM semantics that SLICER represents. SLICER differs structurally from potency-based approaches such as Melanee and MetaDepth in how levels are determined: SLICER derives level membership dynamically from relation types, whereas potency-based approaches assign levels statically through a numeric attribute attached to a clabject, which is decremented at each instantiation, a different kind of rule-based burden that likewise has no counterpart in two-level modelling. Whether the results of this study extend to this class of mechanisms is an open question left for future empirical work.

\paragraph{Conclusion Validity} One threat is that the structural similarity metric of this study treats all relation types equally, whereas in practice some errors have more severe consequences than others; confusing InstN with SpecR, for example, is likely to affect a model's semantic correctness more than other substitutions. This metric does not itself distinguish the severity of errors, but the different types of relation-kind errors have been categorised and discussed individually in RQ3, so the difference in severity is not entirely unrecorded, it is simply distributed across other dimensions of analysis rather than reflected in this single aggregate metric.

A further threat is that judgements of ambiguous cases, such as whether a choice of relation type is ``appropriate,'' may vary between evaluators. This differs from the rater bias discussed in Internal Validity, where evaluators were aware of a model's source; here the concern is one of reliability arising from the subjectivity of the evaluation criteria themselves. We reduce this risk through the quantitative metric definitions and matching rules predefined in Sections~\ref{sec:metrics} and~\ref{sec:Evaluation_Setup}, but for borderline cases at the edge of these rules, disagreement between evaluators has not been eliminated.

A third threat concerns statistical power: each configuration was run only $N=5$ times. We therefore test only factor-level effects and their interaction with non-parametric procedures (Section~\ref{sec:metrics}) rather than differences between individual cells, and effect sizes estimated from samples of this size carry wide uncertainty. Non-significant results, such as the absence of a prompting effect on constraint satisfaction, indicate absence of evidence rather than evidence of equivalence, and the cross-model comparisons in RQ2 are statistically supported only where a test is reported; the remaining observations stay descriptive.
Taken together with the framing given in Section~\ref{sec:metrics}, the statistical tests reported throughout this paper should be understood as corroborating the descriptive findings rather than substituting for them.

\subsection{Future Work}
We foresee the following directions for future work.

\paragraph{Embedding LLMs and AI Agents into the MLM Toolchain}

This study treated the LLM as a stand-alone model generator. The natural next step is to embed LLM assistance into an established MLM toolchain, and our previous non-AI line of work provides the target infrastructure. In earlier work, we combined SLICER-based MLM with multi-view modelling to address horizontal and vertical interoperability in industrial risk management~\cite{DBLP:conf/in4pl/FuGKSS22,DBLP:conf/models/FuGKSS23}. We also contributed the SLICER solution to the Warehouse Challenge, which serves as the reference model of the present study~\cite{DBLP:conf/models/FuSGKS24}. More recently, we extended this line of work into C-SLICER, a collaborative modelling framework that supports both MLM and multi-view modelling~\cite{DBLP:conf/models/FuGKSS25,DBLP:journals/infsof/FuGKSS26}. C-SLICER detects and resolves overlapping and static semantic conflicts, and validates multi-level models against an F-Logic encoding of the SLICER axioms in ErgoAI. The findings of this study map directly onto that infrastructure. First, the logic-based validator can close the generation loop: a generate--validate--repair agent that submits each draft to the ErgoAI checker and feeds the reported axiom violations back to the LLM would provide the precise, rule-level feedback that the generic self-check strategy lacked, targeting exactly the relation-kind errors of Section~\ref{sec:error-patterns}. Second, LLMs can strengthen the semi-automated conflict resolution of C-SLICER by ranking and explaining candidate resolutions instead of presenting them as an undifferentiated list. Third, LLMs offer a natural-language front end for collaborative MLM: stakeholder change requests could be translated into the atomic modification operations on which the conflict management framework already operates, so that AI-generated edits pass through the same detection, resolution, and validation pipeline as manual edits, and, in a longer perspective, autonomous agents could each maintain a view of a shared multi-level model under the same consistency guarantees.

\paragraph{A Benchmark Dataset for Multi-Level Modelling}
The paper's evaluation is limited to a single task and a single language (Section~\ref{sec:threats}), and this limitation cannot be resolved simply by running more experimental trials. The underlying reason is the current absence of systematic, multi-task, multi-language benchmark data for MLM evaluation. In fact, the MLM community has proposed more than one challenge scenario over the years, such as the Bicycle Challenge~\cite{multi2017bicycle}, the Process Challenge~\cite{almeida2019multi}, the Collaborative Comparison Challenge~\cite{multi2021collaborative}, and the Warehouse Challenge~\cite{DBLP:conf/models/KuhneJ23} used in this paper. Each of these challenges has already accumulated reference solutions from different tools and approaches, but this material remains scattered across individual papers and has not yet been organised into a unified, structured dataset directly usable for evaluation. We plan to build on these historical challenges and their published solutions to construct an MLM benchmark dataset covering multiple task scenarios and reference models from multiple MLM languages, providing infrastructure for broader, more statistically meaningful LLM-for-MLM evaluation in future work and addressing the single-scenario limitation of this paper.

\paragraph{LLM for Software Language Engineering}
The evaluation target of this paper, SLICER, is itself a formal modelling language, and the way LLMs complete the task is by being given the language's formal specification in context, rather than relying on knowledge accumulated during pretraining. This setting closely resembles a core class of tasks in software language engineering (SLE): SLE likewise requires LLMs to learn and apply the definition of some formal language (such as a grammar or a metamodel) from context rather than from prior training knowledge, since these languages are similarly low-resource and specialised. Given that this paper has shown LLMs can reliably transform content explicitly stated in the text, while struggling to complete or infer structure that the text leaves implicit, this boundary is equally worth examining in SLE settings. We plan to build on two lines of prior work in this direction. The first concerns DSL style adaptation~\cite{zhang2023creating}. The existing method relies on predefined rules to convert Xtext-generated grammars into a fixed Python style, and we plan to explore having an LLM directly generate a stylistically adapted grammar when given the grammar definition together with examples of a target style, thereby extending the scope of adaptation from a single fixed style to arbitrary, user-defined styles. The second concerns the error reporting mechanism of the EATXT textual editor~\cite{zhang2023technical}. Currently, when a model file's metamodel version is inconsistent with the editor, or when input content does not match the grammar, the editor can only provide a generic message, and we plan to explore feeding structured information about the version difference or grammar mismatch to an LLM so that it can generate a specific, user-facing explanation. Both tasks involve converting explicitly given formal differences or rules into the target output, rather than requiring the model to infer implicit semantics, and therefore fall within the capability boundary identified in this paper, making them relatively low-risk directions for application.


\section{Related Work}
\label{sec:related_work}

This paper sits at the intersection of two lines of research that have not yet converged: one is MLM research that does not involve LLMs, and the other is LLM-based research that remains confined to two-level modelling.

The line of research on MLM itself has accumulated a variety of languages and tools, but all of it relies on human experts to construct or verify models by hand. Fu et al. have carried out a series of works based on the SLICER language: proposing its conceptual framework and applying it to ecosystem and industrial data interoperability scenarios~\cite{DBLP:journals/dke/SelwaySMJGS17},~\cite{DBLP:conf/er/SelwaySMJGS15},~\cite{DBLP:journals/sosym/IgamberdievGSS18}, contributing the SLICER solution to the MULTI Warehouse Challenge (the reference model against which this paper's evaluation is conducted)~\cite{DBLP:conf/models/FuSGKS24}, and further extending it to multi-view collaborative modelling and conflict resolution~\cite{DBLP:conf/in4pl/FuGKSS22},~\cite{DBLP:conf/models/FuGKSS23},~\cite{DBLP:conf/models/FuGKSS25},~\cite{DBLP:journals/infsof/FuGKSS26}. Beyond this work, Atkinson and Kühne were the first to systematically propose the core ideas of multi-level metamodelling and to name the notion of ``accidental complexity''~\cite{DBLP:conf/uml/AtkinsonK01},~\cite{DBLP:journals/sosym/AtkinsonK08}, and de Lara, Guerra, and Cuadrado summarised when and how MLM should be used [18]. Lange et al., addressing the same Warehouse Challenge, contributed an alternative solution using the LML language and the Melanee tool~\cite{lange2023modeling}, providing a natural point of comparison for this paper's SLICER-based solution, and together with Atkinson and Kühne clarified common misconceptions about potency-based deep instantiation~\cite{atkinson2024misconceptions}, which contrasts with the relation-driven approach to level determination used by SLICER. 
Rodríguez and Macías contribute a MultEcore-based solution to the MULTI Process Challenge, a separate community benchmark, illustrating potency-based level determination together with supplementary hierarchies and multilevel coupled model transformations for constraint specification~\cite{rodriguez2019multilevel}.
What these works have in common is that model construction and semantic verification are always carried out by human experts. This paper differs in that it evaluates whether an LLM can, given formal semantic documentation, learn and apply such semantic rules on its own (in this case, those of SLICER).

A second related line of research applies LLMs to modelling tasks for formal target languages that themselves rarely appear in LLM training corpora (even where some of this work is built on mature language workbenches such as Xtext), but the semantic complexity of these languages remains within the scope of two-level modelling. Jiang et al. had LLMs generate transformation code across four model transformation languages, finding that few-shot examples consistently improved syntactic quality while gains in semantic correctness varied by language~\cite{jiang2026llm4mtls}. Zhang et al. used LLMs to automatically adapt Xtext grammars following metamodel evolution, with their method outperforming a rule-based baseline on most languages~\cite{zhang2026b}. The same team also explored using LLMs to support the co-evolution of textual DSL instances with their evolving grammar definitions, first in an exploratory study~\cite{zhang2025leveraging} and later in a systematic evaluation covering ten languages~\cite{zhang2026a}. In addition, Kreikemeyer et al. compared parameter-efficient fine-tuning against few-shot prompting for a formal DSL describing chemical reaction networks, showing that a 7-billion-parameter open-weights model fine-tuned with LoRA could raise its accuracy to a level approaching that of commercial large models, though its target language is essentially a flat set of reaction rules with no notion of abstraction levels or type-instance relationships~\cite{kreikemeyer2025using}. Baumann et al. combined retrieval-augmented generation with few-shot learning for the little-known MontiCore Sequence Diagram DSL, using retrieved example models from a knowledge base to lower the rate of syntactic errors, though their evaluation target remains a conventional two-level modelling artifact~\cite{baumann2024combining}. 
Ben Chaaben et al. propose MAGDA, a tool that uses LLMs and few-shot prompting to assist domain model completion, and evaluate its effect on task completion time, solution elaboration, and user experience through a user study, though it likewise remains confined to conventional two-level domain modelling~\cite{chaaben2026utility}.
Together, these studies show that LLM capability for modelling in this kind of specialised, data-scarce target language has already been reasonably well explored, but the language semantics involved all fall within the scope of two-level modelling and do not touch on the dynamic determination of level membership, the distinction between terminal and extensible instantiation, or the propagation of constraints across levels, all of which are semantic mechanisms specific to MLM. This paper differs in being the first to extend this evaluation paradigm to the semantic level of MLM.

\section{Conclusion}\label{sec:Conclusion}
This paper presents the first systematic empirical investigation of LLM capability on MLM tasks. We had three commercial LLMs (GPT, Claude, and Gemini) generate multi-level models for the MULTI Warehouse Challenge scenario under six prompting strategies, with the full SLICER language semantic documentation provided in context, yielding 90 generated models in total, which we compared against a manually validated reference model using fourteen metrics. The results show that current LLMs can reliably generate valid SLICER models and cover the domain elements and the core instantiation hierarchy, but semantic correctness remains only partial, with recurring errors concentrated in the fine-grained selection of relation kinds and in structure and constraints that the task text implies without stating, indicating that LLMs are better at faithfully transcribing content explicitly stated in the text than at proactively completing implicit structure. Prompting strategies trade off precision against completeness, and self-checking functions mainly as a rule checker. These findings draw a relatively clear line for the division of labour between humans and LLMs in MLM: generating the level structure can be delegated to an LLM, but fine-grained semantic judgements involving relation kinds still require the involvement of a human expert or an automated checking tool.

For future work, one direct direction is to embed LLM generation into an existing MLM toolchain by building a generate–validate–repair loop, feeding the specific violations reported by an axiom checker based on the F-Logic encoding of SLICER's axioms in C-SLICER back to the LLM, in order to provide more precise, rule-level feedback targeting the relation-kind errors identified in this paper. Another direction is to consolidate the multiple challenge scenarios and their published reference solutions already available in the MLM community into a unified, multi-task, multi-language benchmark dataset, to support more statistically meaningful evaluation in future work and to address this paper's limitation of a single scenario and a single language.

\backmatter

\bmhead{Supplementary information}

The replication package accompanying this article is available via the Open Science Framework (OSF)~\cite{FuEtAl2026SoSyMSI}. It contains the reference model, the prompt materials for all six configurations (embedding the condensed task description and the SLICER reference verbatim), all 120 generated models (90 final models and 30 intermediate self-check drafts), the evaluation workbook with the per-model metric measurements, and the figure generation scripts together with the resulting figures.

\bmhead{Acknowledgements}
The work is supported by the Deutsche Forschungsgemeinschaft (DFG, German Research Foundation) - SFB 1608 - 501798263 and supported by the pilot program Core Informatics at KIT (KiKIT) of the Helmholtz Association (HGF).





\bibliography{sn-bibliography}

\end{document}